\documentclass[conference]{IEEEtran}
\IEEEoverridecommandlockouts
\ifCLASSINFOpdf
\else
\fi
\usepackage{amsmath}
\usepackage{diagbox}
\usepackage[english]{babel}
\usepackage{graphicx}
\usepackage{graphicx}
\usepackage{epsfig}
\usepackage{epstopdf}
\usepackage{algorithmic}
\usepackage[linesnumbered,boxed]{algorithm2e}
\usepackage{amsmath}
\usepackage{diagbox}
\usepackage[english]{babel}
\usepackage{graphicx}
\usepackage{graphicx}
\usepackage{epsfig}
\usepackage{epstopdf}
\usepackage{amsthm}
\usepackage{amssymb}
\usepackage{algorithm2e}
\usepackage{subfigure}
\usepackage{hyperref}
\begin{document}
\title{ Asynchronous COMID: The Theoretic Basis for  Transmitted Data Sparsification Tricks on Parameter Server }
\author{
	Cheng Daning$^{1,2}$,Li Shigang$^{1,*}$\thanks{*Corresponding author}, Zhang Yunquan$^1$\\
1.SKL of Computer Architecture, Institute of Computing Technology, CAS, China\\
2.University of Chinese Academy of Sciences\\
Email: \{chengdaning, lishigang, zyq\}@ict.ac.cn
}

\maketitle

\begin{abstract}
Asynchronous FTRL-proximal and $L2$ norm done at server are two widely used tricks in Parameters Server which is a kind of implement of delayed SGD. Their commonness is leaving parts of updating computation on server which reduces the burden of network via making transmitted data sparse. But these two tricks' convergences are not well-proved. In this paper, based on their commonness, we propose a more general algorithm named as asynchronous COMID and prove its convergence. We prove that asynchronous FTRL-proximal and $L2$ norm done at server are applications of asynchronous COMID, which demonstrates the convergences of these two tricks. Then, we conduct experiments to verify theoretical results. Experimental results show that compared with delayed SGD on Parameters Server, asynchronous COMID reduces the burden of the network  without any harm on the mathematical convergence speed and final output. 

\end{abstract}

%%\linenumbers

\section{Introduction}
There are a lot of tricks in machine learning application to get higher training efficiency, better classification accuracy and the ability of solving unconvex optimization. Some of them are reasonable and well-proved, like setting better initial model parameters to reduce training time. But most of other tricks are lack of proof. They can only be used suitably depending on users' experience, like deciding the size of batch and constructing a DNN. In a real situation, the majority of tricks are proved by  experiments instead of rigorous mathematical proofs.

Nowadays, Parameters Server frame, based on delayed SGD algorithms, is the most popular learning frame. However, with the increasing number   of workers, the burden of network would be unaffordable. Asynchronous FTRL-proximal and addressing $L2$ norm on server are two widely used tricks to solve this problem, but they are not rigorously proved. Hereafter, these two tricks will be abbreviated as asynch-FTRL-proximal and $L2$ norm trick. 

These two tricks share the same commonness. They divide updating computation into two parts. One part is computed at worker. The work of this part is scanning dataset, computing the gradient of loss function without regularization term and sending this portion of loss function gradient, a sparse vector, to server. Another part is computed at server. The work of this part is computing the gradient of regularization term and updating model parameters lazily. These two parts are computed asynchronously and separately on servers and workers. Sparse data vectors in first part  reduce the burden of network. 

Based on this commonness, we propose and prove asynchronous Composite Objective MIrror Descent, abbr. asynch-COMID in this paper. Then, we establish the equivalence between asynchronous COMID and the two tricks we mentioned above to prove these two tricks are applications of asynch-COMID. Thus, the convergences of these two tricks are also proved. We fill these gaps between  application and theory of these two tricks via asynch-COMID.

%介绍机器学习和随机优化算法

\subsection{Delayed SGD algorithms and Parameters Server}
SGD, Stochastic Gradient Decent, and Parallel SGD algorithms are one of the hottest topics in machine learning area \cite{Bottou2007The,Shalev2008SVM,Nemirovski2009Robust,Nesterov2009Primal,Dean2012Large,Dekel2012Optimal,Duchi2010Adaptive,chaturapruek2015asynchronous,zhu2016local}.

SGD is designed for following minimization problems
\begin{equation*}
\min c(w )=\frac{1}{m}\sum_{i=1}^{m}{{{c}^{i}}(w)}	
\end{equation*}
where $m$ stands for the amount of sample in dataset, $c^i : \ell_2    \mapsto [ 0,\infty ]$ is convex loss function, and the vector $w \in {{R}^{d}}$.

 $L2$ norm regularized risk minimization is the most widely used loss function, $c^i(w)$ in this case is represented by the following formula:
\begin{equation*}
{{c}^{i}}(w )=\frac{\lambda }{2}{{\left\| w  \right\|}^{2}}+L({{x}^{i}},{{y}^{i}},w \cdot {{x}^{i}})	
\end{equation*}
where $L(\cdot)$ is a convex function in $w \cdot x$.

%%并行SGD的历史，ps架构
Delayed SGD is the most important parallel SGD algorithm.  In delayed SGD algorithm, current model parameters $w_{t}$ adds the gradient of older model parameters in $\tau(t) $ ($\tau(t) < t$) iterations. The iteration step for delayed SGD algorithms is:
\begin{align}
{{w}_{t+1}}&={{w}_{t}}-\eta {{\partial }_{w}}{{c}^{i}}({{w}_{\tau(t)}})\notag
\end{align}
where $\eta$ is the learning rate or step length.

For $L2$ norm regularized risk minimization, the update step is 
\begin{align}
{{w}_{t+1}}=w_{t} -\eta ( L'({{x}^{\tau(t)}},{{y}^{\tau(t)}},w \cdot {{x}^{\tau(t)}}) + \lambda w_{\tau(t)}    ) 	 \label{delay_sgd}		
\end{align}.

Delayed SGD algorithms first appeared in J. Langford's work \cite{Langford2009Slow}. In this work, the $\tau(t)$ function is fixed as Eq. \ref{tau_t}. In Hogwild! Algorithm \cite{Feng2011HOGWILD}, under some restrictions, parallel SGD can be implemented in a lock-free style. Lock-free style means $\tau(t)$ can be any functions which satisfy  $0\leq t - \tau(t) \leq \tau_{max}$.

From the point of view of engineering implementation, the implement of delayed SGD  is Parameters Server. Parameters Server gains high performance via the overlapping the communication time and computation time.  Popular Parameters Server frame includes ps-lite in MXNET \cite{Chen2015MXNet}, TensorFlow \cite{abadi2016tensorflow}, petuum \cite{Xing2013Petuum} and so on. One of the method that constricts the delay was offered by Ho et al \cite{Ho2013More}.

%介绍COMID
\subsection{COMID algorithm and asynch-COMID}
COMID, Composite Objective MIrror Descent, can be treated as a modified SGD. COMID does not linearize regularization term. 
COMID is designed for following regularized loss minimization problem \cite{Duchi2010Composite}.
\begin{equation*}
\min {{c}}(w )=r(w)+\frac{1}{m}\sum_{i=1}^{m}L({{x}^{i}},{{y}^{i}},w \cdot {{x}^{i}})	
\end{equation*}
where $r(w)$ is the convex regularization function like least squares.

The iteration step for COMID is
\begin{equation*}
	w_{t+1}=\mathop{argmin}_{w \in \Omega}\{\eta\left<L_t'(w_t),w-w_t\right>+\eta r(w)+B_\psi(w,w_t) \}
\end{equation*}
where $B_\psi(w,w_t)$ is the Bergman Divergence
\begin{equation*}
B_\psi(w,w_t) = \psi(w) - \psi(w_t) - \left<\nabla \psi(w_t),w-w_t\right>
\end{equation*}
$L_t$ is the abbr. of $L({{x}^{t}},{{y}^{t}},w \cdot {{x}^{t}})$.

In real application, the domain of $w$ is large enough and there exist subgradients $L_t',r'$ in $\partial f,\partial r$. All of these conditions make every $w_t$ satisfy the following optimality condition:
\begin{equation*}
\eta L_t'(w_t) + \eta r'(w_{t+1}) + \nabla \psi(w_{t+1}) - \nabla \psi(w_t) = 0.
\end{equation*}
The diameter of the domain of $w$, i.e. $\Omega$, is $R$, which means the domain of $w$ is large but limited.

%介绍本文的asynch-COMID

In this paper, under more assumptions, we propose following asynchronous COMID iteration steps
\begin{align}
w_{t+1}=&\mathop{argmin}_{w \in \Omega}\{\eta\left<L_{\tau(t)}'(w_{\tau(t)}),w-w_{\tau(t)}\right> \notag  \\
&+\eta r(w)+B_\psi(w,w_t) \} \label{eq1}
\end{align}
where $\tau(t)$ is the delay function, which satisfies $0\leq t - \tau(t) \leq \tau_{max}$.

To make analysis easy, in this paper, we set $\tau(t)$ as
\begin{equation}
\tau(t)=
\begin{cases}
0&{t\leq\tau_{max}}\\
t-\tau_{max}&{t\geq\tau_{max}}
\end{cases}
\label{tau_t}
\end{equation}

The optimality condition of asynch-COMID is
\begin{equation}
\eta L_{\tau(t)}'(w_{\tau(t)}) + \eta r'(w_{t+1}) + \nabla \psi(w_{t+1}) - \nabla \psi(w_t) = 0. \label{eq_opt}
\end{equation}

Asynch-COMID uses the delayed information to update the latest $w$.

In Parameters Server frame, the workers always push delayed information to servers.  When the iteration steps contain delayed information like delayed gradient, the algorithm can run on Parameters Server frame asynchronously.   

In asynch-COMID, part of gradient, $L'(\cdot)$, is delayed information. We can put this part on worker, and other part on server. What is more, the delayed information needs reading sample, but scanning dataset is an exhausting job for computer. When delayed part is calculated on worker, reading dataset time can be hidden by computation and communication time.  This form of asynch-COMID is suitable for running on Parameters Server.

In practice, users can divide the gradient of loss function flexibly to make transmitted data sparse. For example, when $r(w)$ contains $L1$ norm which benefits vector sparsification, it is reasonable to address $L1$ norm on workers.

%介绍FTRL
\subsection{Application 1: Asynch-FTRL-proximal}
With the development of real application, the size of model parameters is extremely large and sparse. SGD is not suitable for this situation. Many sophisticated approaches, such as RDA, FOBOS and so on, do succeed in introducing sparsity. They trade off between accuracy and model parameters' sparsity. COMID is one of the best trade off algorithms. FTRL-Proximal algorithm is the most popular COMID's applications. FTRL-proximal is effective at producing sparse and accuracy model parameters \cite{Mcmahan2013Ad}.

The iteration step of FTRL-Proximal is
\begin{equation*}
w_{t+1}=\mathop{argmin}_w ((L_{1:t}' + \sum_{i=1}^{t-1}r'(w_{i+1}))\cdot w + \widetilde{\psi}_{1:t}(w)+r(w)) 
\end{equation*}
where $L'_{1:t}  $ is the short hand for $\sum_{i=1}^{t}L'_t(w_t)$, $\psi_t$ be a sequence of differentiable origin-centred convex functions ($\nabla \psi_t(0) = 0$) and $\widetilde{\psi_t}(w) = \psi_t(w-\hat{w_t})$.

There are two versions widely used asynchronous FTRL-proximal.
The first version is 
 \begin{align*}
&w_{t+1}=\mathop{argmin}_w ((L_{\tau(1):\tau(t)}' + \sum_{i=1}^{t-1}r'(w_{i+1}))\cdot w\\
& + \widetilde{\psi}_{\tau(1):\tau(t)}(w)+r(w))
\end{align*}
The second one is :
\begin{equation*}
w_{t+1}=\mathop{argmin}_w ((L_{\tau(1):\tau(t)}' + \sum_{i=1}^{t-1}r'(w_{i+1}))\cdot w + \widetilde{\psi}_{1:t}(w)+r(w)) 
\end{equation*}
In this paper,  we only discuss second FTRL-proximal. In following sections, asynchronous FTRL-proximal means the second version. We show equivalence between asynchronous FTRL-proximal and asynchronous COMID.

\subsection{Application 2: $L2$ norm trick}
%介绍L2范式

$L1$, $L2$ norm are the most widely used regularization methods. $L1$ norm is mainly used to produce sparse solution. $L2$ norm, ridge regression, is the most commonly used method of regularization of ill-posed problems.

\begin{table}[!thbp]
	\centering
	\begin{tabular}{| m{1.7cm}| m{1.7cm}| m{1.7cm}| m{1.7cm}|}
		\hline
		Dataset & Source &Number of features in  a sample & Number of none zero features in a sample\\
		\hline
		KDD 2010(algebra) &KDD CUP 2010& 20216830 & 20-60 \\
		\hline
		 Avazu & Avazu's Click-through Prediction & 1,953,951 & 30-60 \\
		\hline
		Minist8m& MNIST & 780 & 130-200\\
		\hline
		 Webspam&webb spam corpus&16609143& 70 -90\\ 
		\hline
		KDD 2012& KDD CUP 2012&54686452& 10-40\\
		\hline
	\end{tabular}
		\caption{\label{table_1}different Datasets and its sparseness}
\end{table}

Using normal Parameters Server method, i.e. Eq. \ref{delay_sgd},  workers should send the gradients to server. The $L2$ norm should be a part of loss function as theoretical analysis mentioned \cite{Feng2011HOGWILD,Langford2009Slow,Zinkevich2010Parallelized}.

Most of the time, $L'(\cdot)$  is sparse vector. The sparsity of $L'(\cdot)$ often corresponds to the sparsity of sample vector, like the cases of linear classifier and fully connected neural network. Table \ref{table_1} shows the sparsity of sample in different datasets.

However, $L2$ norm exerts great press on network for the gradient of $L2$ norm in loss function is a dense vector. Basically, gradient of $L2$ norm  is the product of model parameters and a constant number. When using normal method, workers have to send a dense vector in network which would be  a heavy burden for network. Especially nowadays, the number of features in sample is extremely large. 

There is a trend in real application that when training model parameters, the coders often get rid of $L2$ norm to gain high performance. It is a trade off between training efficiency and classification accuracy. Another method to deal with this problem is $L2$ norm trick. In $L2$ norm trick, the computation of $L2$ norm is done at server, using the latest model parameters. The burden of network is reduced. What is more, $L2$ norm trick is suitable for lazy updating. Many Parameters Server frames use this kind of method to deal with $L2$ norm like PaddlePaddle\cite{Paddle2016}, but none of them shows its reasonability. The iteration step of $L2$ norm trick is described as follows:
 \begin{equation*}
 w_{t+1} =  w_t -\eta(L_{\tau(t)}(w_{\tau(t)}) + \lambda w_t)  
 \end{equation*}

In our work,  we establish the equivalence between our asynchronous COMID and $L2$ norm trick. Thus, we proved that addressing $L2$ norm on server would not harm algorithm convergence.

\subsection{Summary}
The key contributions of this paper are as follows:

1. We offer the proof of asynchronous COMID. Asynchronous COMID can work on Parameters Server frame.

2. Based on the asynchronous COMID, we prove that it is reasonable that FTRL-Proximal algorithm runs on Parameters Server frame asynchronously. We also conduct experiments to verify this theoretical result.

3. Based on the asynchronous COMID, we prove that $L2$ norm trick is reasonable. We also conduct experiments to verify this theoretical result.

In Section 2, we will demonstrate the proof details and theoretical results. In Section 3, we will present the experimental results.

\section{Proof and analysis}
%%假设与前提
\subsection{Notations, setting and assumptions}
Before continuing, except above mentioned notations, we establish more notations in this subsection. The subdifferential set of a function $f$ evaluated at $w$ is denoted $\partial f$ and a particular subgradient by $f'(w) \in \partial f(w)$. When a function is differentiable, we write $\nabla f(w)$. The inner product for $u,v$ is $\left<u,v\right>$ or $u \cdot v$. We focus on regularized loss function, in which the goal is to achieve low regret w.r.t. a static predictor $w^* \in \Omega$ on a sequence of functions $c^t(w) = L_t(w) +r(w)$. Here $L_t$ and $r$ ($r > 0$) are convex functions in $\Omega$, and $\Omega$ is convex set. We seek bounds on the regularized regret with respect to the minimum $w^*$, defined as
\begin{align*}
regret_c(T,w^*) &\triangleq \sum_{t=1}^{T}[c^t(w_t)-c^t(w^*)]\\
&=\sum_{t=1}^{T}[L_t(w_t)+r(w_t)-L_t(w^*)-r(w^*)]
\end{align*}

Throughout, $\psi$ designates a continuously differentiable function that is $\lambda$-strongly convex w.r.t. a norm $\left\| \cdot \right\| $ on set $\Omega$,
\begin{equation}
B_{\psi}(w,v) \geq \lambda/2\left\| w-v\right\|^2 \label{eq3}%% \lable{eq3}
\end{equation}
In this paper, we also assume $\psi$ satisfies following inequation.
\begin{equation}
\alpha \left\| \nabla \psi(w) - \nabla \psi(v) \right\| \geq  \left\| w - v\right\| \label{eq4} %%\lable{eq4}
\end{equation}
where $\alpha$ is a constant.

In the proof of asynchronous algorithm, it is usual to limit the  norm of gradient as following equations. We also have to obey this limit. In some works, these limitations are on whole loss functions, which contain regularization part  \cite{Feng2011HOGWILD,Langford2009Slow,Zinkevich2010Parallelized}.
\begin{equation}
\left\| c'_t(w) \right\|  \leq M_{out} \label{eq5}
\end{equation}
almost sure for all $w \in \Omega$.

And for some works, like Composite Objective MIrror Descent \cite{Duchi2010Composite}, the limitation is just on $L(\cdot)$. This limitation is presented as follows
\begin{equation}
\left\| L'_t(w) \right\|  \leq M_{in}  \label{eq6}
\end{equation}

\subsection{Asynchronous COMID}
%%异步COMID算法
Our proof is based on original COMID proof \cite{Duchi2010Composite}. We use Eq.\ref{eq5}, \ref{eq6} to bound the "progress bound". Following lemma is the base of later proof.

\textbf{Lemma 1} Let the sequence $\{w_t\}$ be defined by the update in Eq. \ref{eq1}. Under the limitation of Eq. \ref{eq3} \ref{eq4} \ref{eq5} \ref{eq6}. For any $w \in \Omega$,
\begin{align}
&\eta(L_{\tau(t)}(w_{\tau(t)}) - L_{\tau(t)}(w*) + r(w_{t+1}) - r(w^*) )\notag\\
&\leq B_{\psi}(w^*,w_t) - B_{\psi}(w^*,w_{t+1})\notag\\
&+\eta^2\alpha\tau_{max}(2M_{in}^2+M_{in}M_{out})
\end{align}
\begin{proof}
	We have
	\begin{align*}
		&\eta\left[L_{\tau(t)}(w_{\tau(t)})+r(w_{t+1})-L_{\tau(t)}(w^*)-r(w^*)\right]\\
		&\leq \eta\bigg[\left<w_{\tau(t)}-w^*,L_{\tau(t)}'(w_\tau(t))\right> \\
		&+\left<w_{t+1}-w^*,r'(w_{t+1})\right>\bigg]\\
		&=\eta\bigg[\left<w_{t+1}-w^*,L_{\tau(t)}'(w_\tau(t))\right>\\
		&+\left<w_{t+1}-w^*,r'(w_{t+1})\right>\bigg]\\
		&+\eta\left< w_{\tau(t)} - w_{t+1},L_{\tau(t)}'   (w_{\tau(t)})             \right>\\
	\end{align*}
	Using the optimality condition, i.e. Eq.\ref{eq_opt}
	\begin{align*}
		&=\bigg<w^*-w_{t+1},\nabla \psi(w_t)-\nabla\psi(w_{t+1})-\\
		&\eta L_{\tau(t)}'(w_{\tau(t)})-\eta r'(w_{t+1})\bigg>\\
		&+\left<w^*-w_{t+1},\nabla \psi(w_{t+1}) -\nabla \psi(w_t)\right>\\
		& + \eta\left< w_{\tau(t)} - w_{t+1},L_{\tau(t)} '   (w_{\tau(t)})             \right>\\
		&=B_{\psi}(w^*,w_t)-B_{\psi}(w_t,w_{t+1})-B_{\psi}(w^*,w_{t+1})\\
		&+\eta\left< w_{\tau(t)} - w_{t+1},L_{\tau(t)}'    (w_{\tau(t)})\right>  \\
	\end{align*}
	Noting that Bregman divergences are always non-negative and using Eq. \ref{eq4}
	\begin{align*}
		&\leq B_{\psi}(w^*,w_t)-B_{\psi}(w^*,w_{t+1})\\
		&+\eta \left\|  w_{\tau(t)} - w_{t+1}\right\| \left\| L_{\tau(t)}  '  (w_{\tau(t)}) \right\|  \\
		&\leq B_{\psi}(w^*,w_t)-B_{\psi}(w^*,w_{t+1})\\
		&+\eta\alpha\left\| \nabla \psi( w_{\tau(t)}) -\nabla \psi( w_{t+1})\right\| \left\| L_{\tau(t)}  '  (w_{\tau(t)}) \right\|	\\
		&\leq 	B_{\psi}(w^*,w_t)-B_{\psi}(w^*,w_{t+1})\\&+\eta\alpha \sum_{i = \tau(t)}^{t}\left\| \nabla \psi( w_{i}) -\nabla \psi( w_{i+1})\right\| \left\| L_{\tau(t)}  '  (w_{\tau(t)}) \right\|
	\end{align*}
	Using the optimality condition (Eq.\ref{eq_opt})  and lmitations ( Eq. \ref{eq5} \ref{eq6} )
	\begin{align*}
		&\leq 	B_{\psi}(w^*,w_t)-B_{\psi}(w^*,w_{t+1})\\
		&+\eta^2\alpha \sum_{i = \tau(t)}^{t}\left\| L_{\tau(i)}'(w_{\tau(i)})+r'(w_{i+1})\right\| \left\| L_{\tau(t)}  '  (w_{\tau(t)}) \right\|\\
		&=	B_{\psi}(w^*,w_t)-B_{\psi}(w^*,w_{t+1})+\\
		&\eta^2\alpha \sum_{i = \tau(t)}^{t}\left\| L_{\tau(i)}'(w_{i})+c'(w_{i+1})-L_{i+1}'(w_{i+1})\right\| \cdot\\
		& \left\| L_{\tau(t)}  '  (w_{\tau(t)}) \right\|\\
		&\leq 	B_{\psi}(w^*,w_t)-B_{\psi}(w^*,w_{t+1})\\
		&+\eta^2\alpha(t-\tau(t))(2M_{in}^2+M_{in}M_{out})\\
		&\leq 	B_{\psi}(w^*,w_t)-B_{\psi}(w^*,w_{t+1}) \\
		&+\eta^2\alpha\tau_{max}(2M_{in}^2+M_{in}M_{out})		
	\end{align*}
\end{proof}

The following theorem uses Lemma 1 to establish a general regret bound for the COMID framework.

\textbf{Theorem 2} Let the sequence ${w_t}$ be defined by the update in Eq. \ref{eq1}. Then for any $w^*\in \Omega$
\begin{align}
&regret_c(T,w^*)\leq\frac{1}{\eta}B_{\psi}(w^*,w_1) \notag\\
&+\sum_{i=1}^{\tau_{max}}r(w_i)+\tau_{max}\eta\alpha T(2M_{out}^2+M_{in}M_{out})
\end{align}
\begin{proof}
	By Lemma 1,
	\begin{align*}
	&\eta\sum_{t=1}^{T}(L_t(w_t)-L_t(w^*)+r(w_{t+\tau})-r(w^*))\\
	&\leq B_{\psi}(w^*,w_1)-B_{\psi}(w^*,w_{T+1})\\
	&+\tau_{max}\eta^2\alpha T(2M_{out}^2+M_{in}M_{out})
	\end{align*}
	Noting that Bregman divergences are always non-negative, and our assumption that $r(w)>0$. Adding $\sum_{i=1}^{\tau_{max}}r(w_i)$ to both sides of the above mentioned equation and dropping the $\sum_{i=T}^{T+\tau_{max}}r(w_i)$. Then, we get
	\begin{align*}
	&regret_c(T,w^*)\leq\frac{1}{\eta}B_{\psi}(w^*,w_1) \notag\\
	&+\sum_{i=1}^{\tau_{max}}r(w_i)+\tau_{max}\eta\alpha T(2M_{out}^2+M_{in}M_{out})
	\end{align*}
\end{proof}
In fact, there is no need to require what $\tau(t)$ is. It is obvious that $regret_c(T,w^*)\leq\ constant_1 + \eta T constant_2$, if $\tau(t)$ is a  function which is almost surjective to $\mathbb{N}$ with finite elements' missing and duplication. We use Eq. \ref{tau_t} just because it is easy to present our main idea.

For $regret_c(w^*,T)$ is the Cesaro Sum of $c_t(w_t)-c_t(w^*)$, the $c_t(w_t) - c_t(w*)$ is convergence to 0, when $\eta$ is small enough.

\subsection{Equivalence between asynchronous FTRL-proximal and asynchronous COMID}
%%PS-FTRL算法
Before our proof, we introduce a lemma by H.Brendan McMahan without proof .

\textbf{Lemma 3\label{Lemma3}} Let $F:\mathbb{R}^n \mapsto \mathbb{R}$ be strongly convex with continuous partial derivatives, and let $\Phi :\mathbb{R}^n \mapsto \mathbb{R}$ be an arbitrary convex function. Define $g(x) = F(x) + \Phi(x)$. Then, there exists a unique pair $\left<x^*,\phi\right>$ such that both
\begin{equation*}
\phi' \in \partial \Phi(x^*)
\end{equation*}
and
\begin{equation*}
 x^*=\mathop{argmin}_x F(x)+\phi' \cdot x
\end{equation*}
Further, this $x^*$ is the unique minimizer of $g(x)$

Noting that an equivalent condition to $x^* =\mathop{argmin}\limits_{x}( F(x) + \phi'\cdot x)$ is $\nabla F(x^*) + \phi' =0$

Lemma 3 shows that there exists a sub-gradient which satisfies
\begin{equation}
\mathop{argmin}_x F(x)+\Phi(x)=\mathop{argmin}_x F(x)+\phi' \cdot x
\end{equation}\cite{Mcmahan2013Follow}.

\textbf{Theorem 4} Let $\psi_t$ be a sequence of differentiable origin-centred convex functions ($\nabla \psi_t(0) = 0$), with $\psi$ strongly convex. Let $w_0=\widetilde{w}_0 = 0$. For a sequence of loss functions $L_t(w) = g_t\cdot w +r(w)$, let the sequence of points $\widetilde{w}_t$ played by asynchronous COMID be
\begin{equation}
\hat{w}_{t+1} = \mathop{argmin}_w ( L_{\tau(t)}' \cdot w +r(w) + \widetilde{B}{_{1:t}(w,\hat{w_t})}) \label{COMID2FTRL}
\end{equation}
where $\widetilde{\psi_t}(w) = \psi_t(w-\hat{w_t})$, and $\widetilde{B}_t = B_{\widetilde{\psi}_t}$, so $\widetilde{B}_{1:t}$ is the Bregman divergence with respect to $\sum_{i=1}^{t}\widetilde{\psi}_i$. Consider the alternative sequence of point $w_t$ played by a proximal FTRL algorithm, applied to those same $L_t$, defined by
\begin{equation}
w_{t+1}=\mathop{argmin}_w ((L_{\tau(1):\tau(t)}' + \sum_{i=1}^{t-1}r'(w_{i+1}))\cdot w + \widetilde{\psi}_{1:t}(w)+r(w)) \label{FTRL2COMID}
\end{equation}
Then, these algorithms are equivalent, in that $w_t = \hat{w}_t$ for all $t\geq 0$.

\begin{proof}
	The proof is by induction. For the base case, we have $w_0$=$\hat{w}_0$. From the  optimality condition and Lemma 3 we know that there exists a unique $r'(w_t)\in \partial r(w_t)$
	\begin{equation*}
	L_{\tau(1):\tau(t-1) }' + \sum_{i=1}^{t-2}r'(\hat{w}_{i+1})+  \nabla\widetilde{\psi}_{1:t-1}(\hat{w}_{t})+r'(\hat{w}_{t}) = 0
	\end{equation*}
	then
	\begin{equation}
	-\nabla \widetilde{\psi}_{1:t-1}(\hat{w}_t) = L_{\tau(1):\tau(t-1) }' +\sum_{i=1}^{t-1}r'(\hat{w}_{i+1}) \label{FTRL2COMID_middle}
	\end{equation}
	Then, starting from Eq. \ref{COMID2FTRL}
	\begin{equation*}
	\hat{w}_{t+1} = \mathop{argmin}_w ( L_{\tau(t)}' \cdot w +r(w) + \widetilde{B}{_{1:t}(w,\hat{w_t})})
	\end{equation*}
	Using Lemma 3
	\begin{equation*}
	\hat{w}_{t+1} = \mathop{argmin}_w ( L_{\tau(t)}' \cdot w +r'(\hat{w}_{t+1})w + \widetilde{B}{_{1:t}(w,\hat{w_t})})
	\end{equation*}	
	Using the definition of $\widetilde{B}{_{1:t}(w,\hat{w_t})}$
	\begin{align*}
	 &\hat{w}_{t+1}=\mathop{argmin}_w ( L_{\tau(t)}' \cdot w +r'(\hat w_{t+1})w + \widetilde{B}{_{1:t}(w,\hat{w_t})})\\
	 &=\mathop{argmin}_w( L_{\tau(t)}'\cdot w + \widetilde{\psi}_{1:t}(w)-\widetilde{\psi}_{1:t}(\hat w_t)\\ &-\nabla\widetilde{\psi}_{1:t}(\hat w_t)(w-w_t)   +r'(\hat w_{t+1})w  )
	\end{align*}
	Dropping the term independent of $w$ and $\nabla\widetilde{\psi}_{t}(w_t) = 0$
	\begin{align*}
	&=\mathop{argmin}_w( L_{\tau(t)}'\cdot w + \widetilde{\psi}_{1:t}(w)-\nabla\widetilde{\psi}_{1:t}(\hat w_t)w  \\ 
	& +r'(\hat w_{t+1})w  )\\
	&=\mathop{argmin}_w( L_{\tau(t)}'\cdot w + \widetilde{\psi}_{1:t}(w)-\nabla\widetilde{\psi}_{1:t-1}(\hat w_t)w \\ 
	&  +r'(\hat w_{t+1})w  )	
	\end{align*}
	Using Eq.\ref{FTRL2COMID_middle}, we get
	\begin{equation*}
	w_{t+1}=\mathop{argmin}_w ((L_{\tau(1):\tau(t)}' + \sum_{i=1}^{t-1}r'(w_{i+1}))\cdot w + \widetilde{\psi}_{1:t}(w)+r(\hat w))
	\end{equation*}	
\end{proof}

\subsection{Equivalence between asynchronous COMID and $L2$ norm trick}
%%L2范式证明2
When $r(w) = \lambda/2\left\| w\right\|^2 $, the $c_t(w) = L_t(w) + r(w)$ is the $L2$ norm regularization loss function. Here, we use $\lambda$ because if the loss function is  $c_t(w) = L_t(w) + \lambda/2\left\| w\right\|^2$ and  $L_t(w)$ is convex function, the $c_t(w)$ is at least $\lambda$-strongly convex function, like $L2$ norm regularization hinge loss for SVM.

Explicit asynchronous COMID algorithm with $\psi(w) =  1/2 \left\| w\right\|^2 $ is as follows
\begin{equation}
w_{t+1} = \frac{1}{1+\lambda\eta} w_t -\eta L_{\tau(t)}(w_{\tau(t)}) \label{expl-comid}
\end{equation}
When $\lambda\eta$ is small, using Taylor expansion, Eq. \ref{expl-comid} is the same as
 \begin{equation}
 w_{t+1} =  w_t -\eta(L_{\tau(t)}(w_{\tau(t)}) + \lambda w_t) \label{half-delay}
 \end{equation}
Comparing with Eq. \ref{delay_sgd}, we can find that Eq.\ref{half-delay} can put their regularization term on server in Parameters Server.

\section{Numerical experiment}
\subsection{Platform}
Our experiments are conduced on Era supercomputer which consists of Xeon E5-2600v3 2.6G CPU connected by Infiniband.
\subsection{Dataset}
We use the data from Avazu's Click-through Prediction as our experiment data. Dataset is used in competition on click-through rate prediction jointly hosted by Avazu and Kaggle in 2014. We use part of the winning solution version data from Yuchin Juan et al. \cite{Yu2010Feature}, named as avazu-site.tr. Each sample in this dataset has 1000000 features.
\subsection{Evaluation}
For the evaluation criterion, we use the $logloss$ as $L(\cdot)$. In presentation, we use  logloss of dataset. The logistic loss of dataset is defined as
\begin{equation*}
logloss_{dataset}=\sum_{i=1}^{m}log(1+exp(-y^i \cdot(w\cdot x^i)))
\end{equation*}
where $m$ is the size of dataset. To clearly show the gap, we will adjust the size of test dataset in different experiments. In following part, all $logloss$ is the $logloss_{dataset}$.

\subsection{Implement}
Our implements of above mentioned algorithms are basic version of optimization algorithms. The batch size is 1 for all implements. Our implements do not include any additional terms like bias term because our goal of following experiments is to show those tricks do not harm final output and convergence speed instead of seeking better model parameters which correspond to lower logistic loss.

\subsection{Asynch-FTRL-proximal experiments setting and result}

\begin{algorithm}[!htp]
	\KwIn{Parameters $\alpha$, $\beta$, $\lambda_1$, $\lambda_2$}	
	\SetAlgoLined
	\SetKwFunction{worker}{worker}
	\SetKwFunction{server}{server}
	\SetKwProg{myalg}{For Worker:}{}{}
	\myalg{}{
		\For{$t_1 = 1$ $\to$  Forever}{
			Pick Sample $x_{t_1}$ and its label $y_{t_1}$\;
			Pull the latest model parameters $w_t$ from Server\;
			Calculate $p_{t_1} = sigmod(w_{t_1} \cdot x_{t_1}) $\;
			Calculate $L_{t_1} = (p_{t_1} - y_{t_1})x_{t_1}$ \;
			Push $L_{t_1}$\;
		}
	}{}
	%\KwRet\\
	\SetKwProg{myalg}{For Server:}{}{}
	\myalg{}{
		Initialize $z$ as requirement. 
		Initialize every feature of $n$ as 0\;
		\For{$t_2$ = 1 $\to$ Forever }{
			Receive $L_{\tau(t_2)}$ from one of workers\;
			Let $I \in \{ i\mbox{ } | \mbox{ the } i\mbox{th feature in }   L_{\tau(t_2)} \neq 0\}$\;
			\For{ $i \in I$}{
				When ${\mbox{if } \left| z_{t_2,i}\right| \leq \lambda }$\\
				$w_{t_2,i}= 0$\;
				Else\\
				$w_{t_2,i}= -(\frac{\beta + \sqrt{n_{t_2,i}}}{\alpha} + \lambda_2)^{-1}(z_{t_2,i} - sgn(z_{t_2,i}) \lambda_1)$\;
			}
			\For{$i \in I$}{
				$\sigma_{t_2+1,i} = \frac{1}{\alpha}(\sqrt{n_{t_2,i} + L_{\tau(t_2)}^2} - \sqrt{ n_{t_2,i} }         )  $\;
				$z_{t_2+1,i} = z_{t_2,i} + L_{\tau(t_2)} - \sigma_{t_2,i}w_{t_2,i}$\;
				$n_{t_2+1,i} = n_{t_2,i} + L_{\tau(t_2),i}^2$		
			} 
			\For{$i \notin I$}{
				$\#$ this part does really work
				$\sigma_{t_2+1,i} = \sigma_{t_2,i}  $\;
				$z_{t_2+1,i} = z_{t_2,i} $\;
				$n_{t_2+1,i} =n_{t_2,i}$		
			} 	
		}
	}
	%\KwRet
	\caption{Per-Coordinate asynch-FTRL-Proximal with $L_1$ and $L_2$ Regularization for Logistic Regression on Parameters Server}
	\label{FTRL on PS}
\end{algorithm} 

\begin{figure*}[!htp]
	
	\subfigure[ ]{
		
		\includegraphics[width=0.5\textwidth]{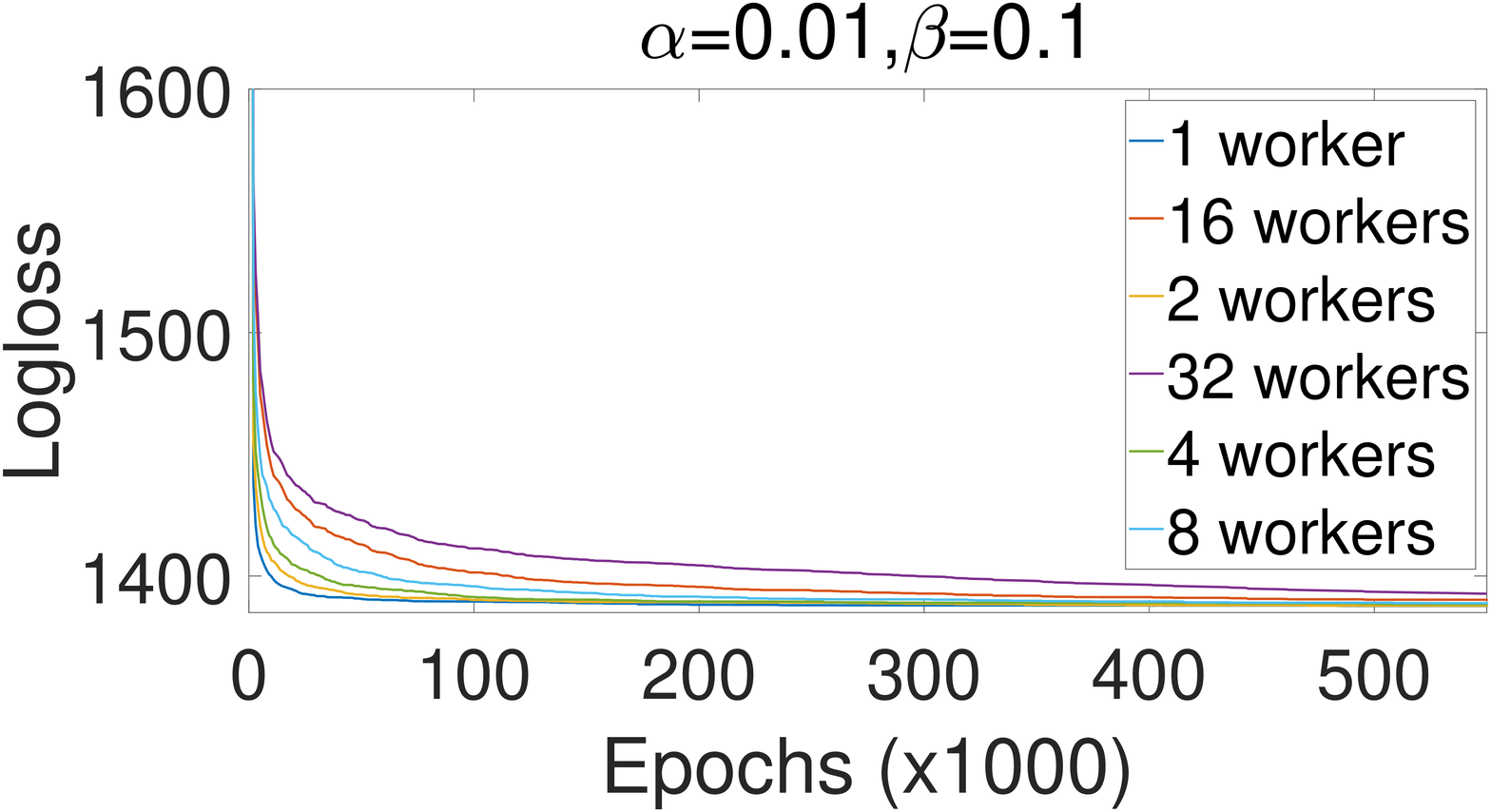}
		
	}
	\subfigure[ ]{

		\includegraphics[width=0.5\textwidth]{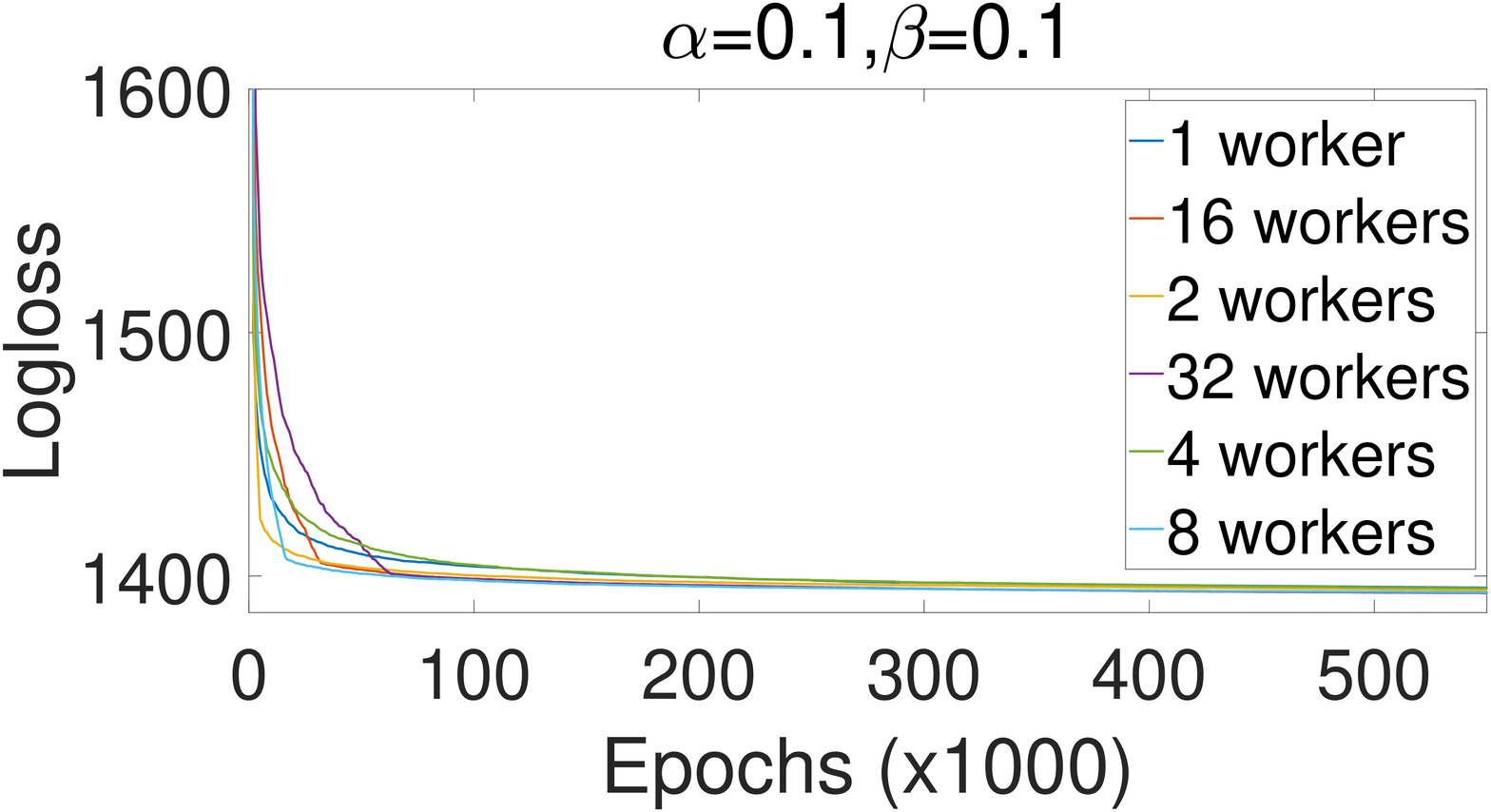}

	}
	\subfigure[ ]{
		
		\includegraphics[width=0.5\textwidth]{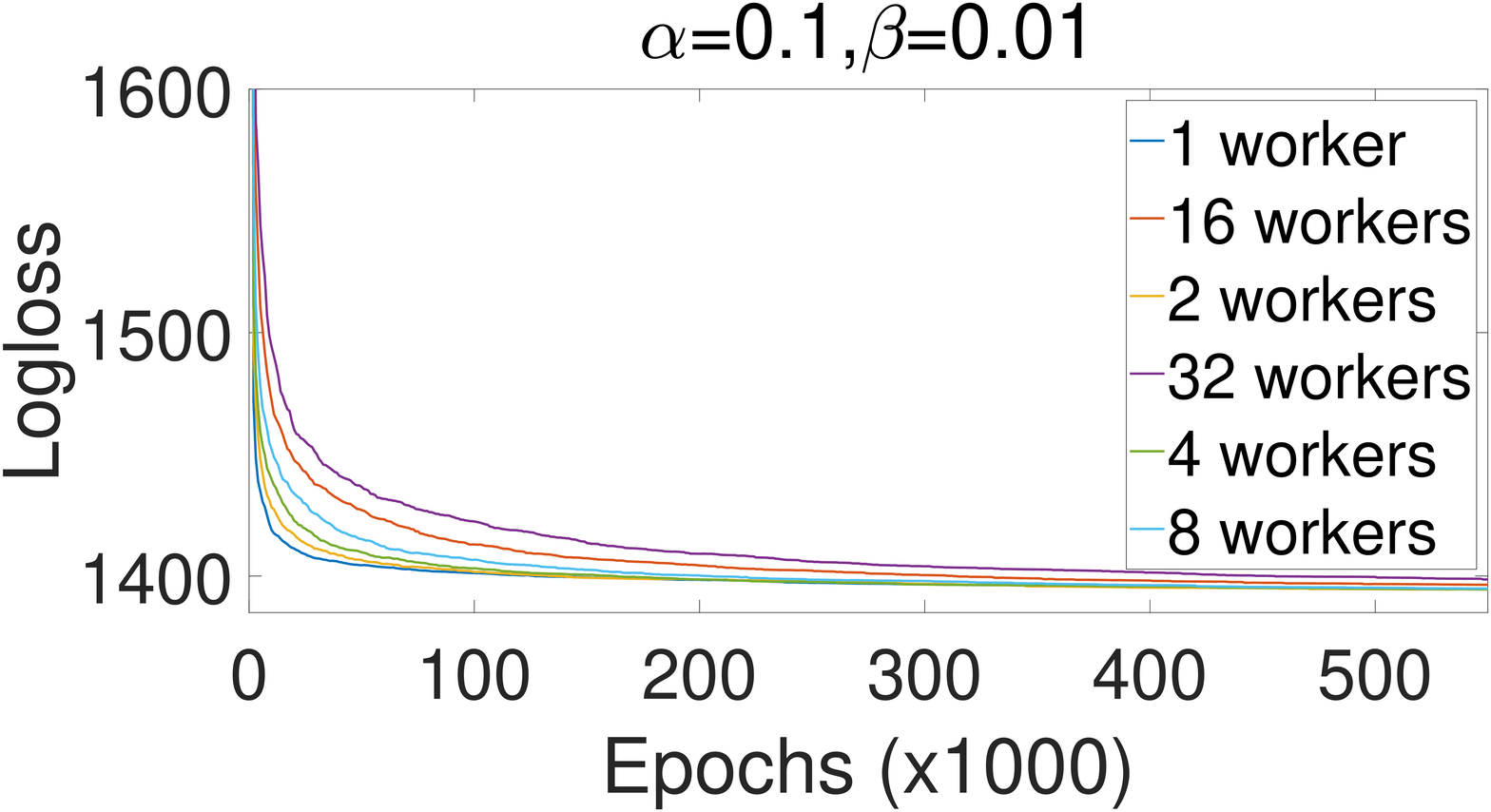}

	}
	\subfigure[ ]{

		\includegraphics[width=0.5\textwidth]{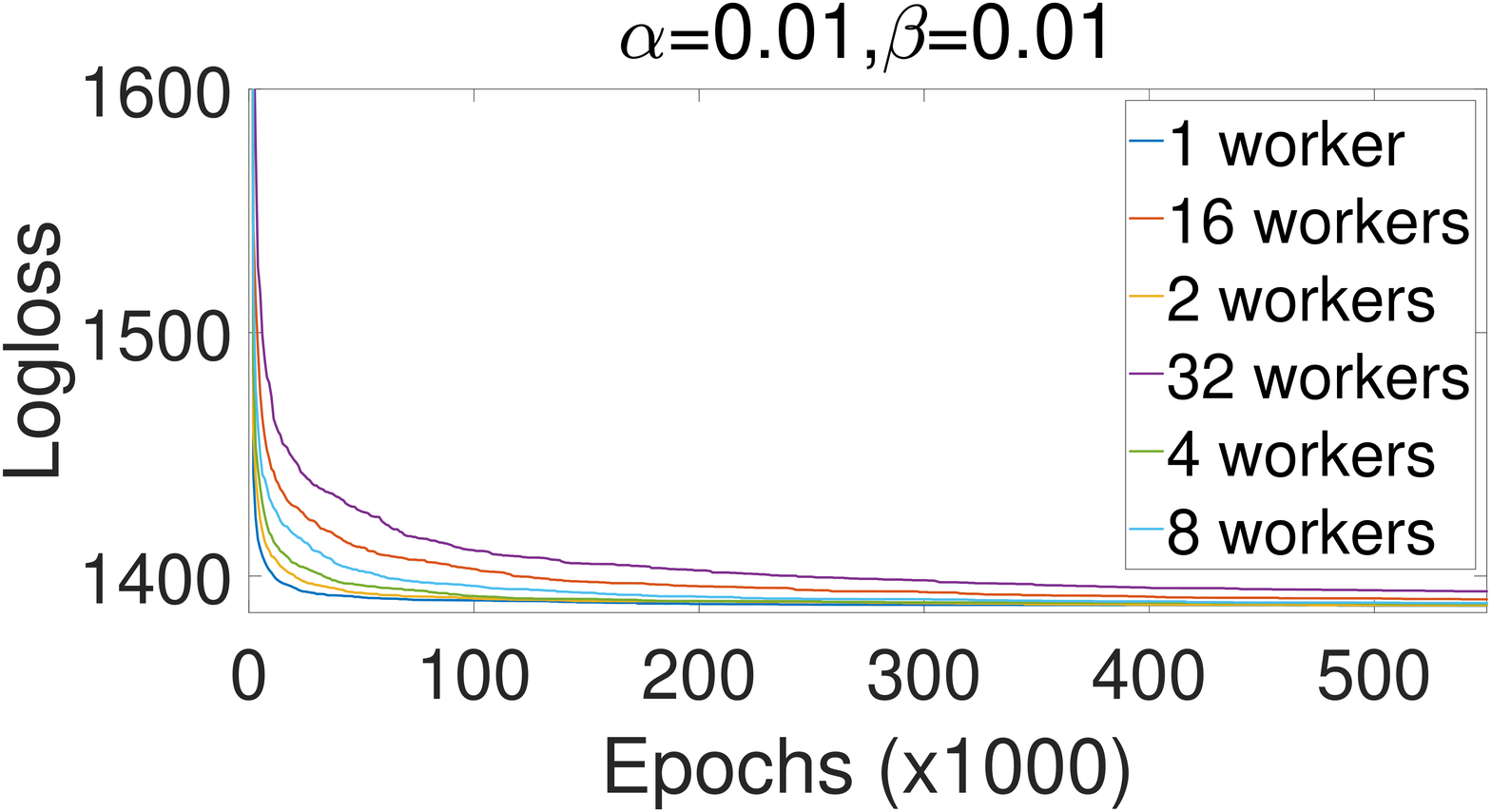}
		
	}
	\caption{the performance of asynch-FTRL-proximal with different number of worker and parameter}	
	\label{different_thread_FTRL}
\end{figure*}

In this experiment, we will show the gaps which are between sequential FTRL-proximal algorithm and asynch-FTRL-proximal on a Parameter Server platform.  The asynch-FTRL-proximal with logloss algorithm implement is described as algorithm \ref{FTRL on PS}.

When the number of worker is 1, asynchronous FTRL-proximal would degenerate into a sequential FTRL-proximal.

\textbf{Seting} We use the test dataset which contains 2700 samples. We set $\lambda_1 = 0.01$ and $\lambda_2 = 0.001$. Because in this dataset, the minimum of logistic loss is close to zero, we have to adjust $z_1$ to let $w_1 = (1,1,\cdots,1)$ to have more number of epochs.

We conduct 4 experiments in all. In each experiment, we fix the value of $(\alpha,\beta)$ and change the number of workers.

\textbf{Result} Figure \ref{different_thread_FTRL} shows the result of 4 experiments. All of those experiments prove that under different parameter setting, the convergence speed would slow with the increasing number of workers. But the gaps between different curves are small. All of those experimental results also present that asynch-FTRL-proximal does not harm final output.

\subsection{$L2$ norm trick experiments setting and result}
In this experiment, we will show the gaps which are between the $L2$ norm trick and the normal method where transmitted data  contains $L2$ norm  on Parameters Server.  Our goal is to show that these gaps are small, which means that the convergence speed of $L2$ norm on server trick is closed to traditional method without any tricks.

\begin{figure}[!htp]
	\centering
	\includegraphics[height = 4cm, width = 8 cm]{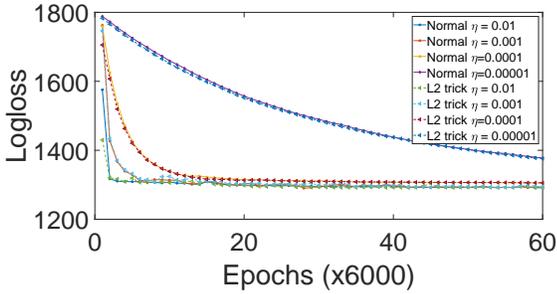}
	\caption{Impact of $\eta$ :  The gap between traditional method and trick method in different setting}
	\label {different_eta_L2norm}
\end{figure}

\begin{figure}[!htp]
	\centering
	\includegraphics[height = 4cm, width = 8 cm]{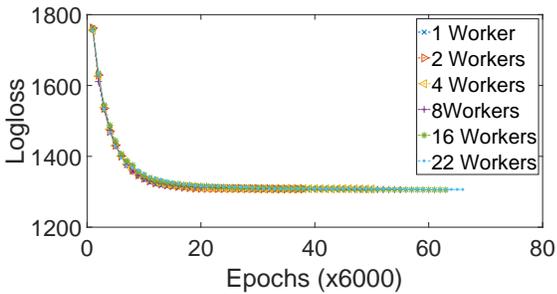}
	\caption{Impact of multi-worker:  The convergence speed of using different number of workers, with $\eta = 0.0001, \lambda = 0.001$}
	\label {different_thread_L2norm}
\end{figure}

\textbf{Setting} We use the test dataset which contains 2500 samples. The initial value is 0 for all features in model parameters.

In impact of $\eta$ experiment, we set $\lambda = 0.001$ and use 10 workers. Via different curves with various $\eta$, we can see the sensibility of $L2$ norm trick for different $\eta$.

In impact of multi-workers experiment, we set $\lambda = 0.001, \eta = 0.001$. Via different curves with various number of workers, we can see the sensibility of $L2$ norm trick for multi-workers.

\textbf{Result}
Figure \ref{different_eta_L2norm} shows the performance of normal method and trick method in different $\eta$ setting.  With  $\eta= 0.01, 0.001,0.0001,0.0001$, the gaps between two methods are small. These phenomenons present that $L2$ norm trick does not harm convergence speed and final output under different parameters setting.

Figure \ref{different_thread_L2norm} shows the performance of normal method and trick method in different number of workers setting. As we can see, when changing the number of worker, $L2$ norm trick does not harm convergence speed and final output.

\section{Conclusion and future work}
In this paper, we propose and prove the convergence of asynch-COMID algorithm.  Asynch-COMID  reduces the burden of network by making transmitted data sparse. We prove that two widely used tricks, $L2$ norm trick and asynch-FTRL-proximal, are applications of asynch-COMID. We also demonstrate that for certain kinds of dataset, $L2$ norm trick and asynch-FTRL-proximal exert tiny influence on convergence speed and final output.

For the future work, we will discuss more mathematical properties of asynch-COMID besides $regret$. What is more, we want to investigate the  mathematical properties of dataset and loss function which determine the gap of convergence speeds and distance of the outputs from different training algorithm. It is also interesting to offer the proofs of more unproved tricks.

\section{Acknowledgement}
This work was supported by the National Natural Science
Foundation of China under Grant No. 61432018, Grant No.
61502450, Grant No. 61521092, and Grant No. 61272136;
National Major Research High Performance Computing
Program of China under Grant No. 2016YFB0200800.

\bibliography{IEEEabrv,mybibfile}
\bibliographystyle{IEEEtran}
%\bibliographystyle{aaai}

% that's all folks
\end{document}